\documentclass[%
reprint,
superscriptaddress,
 amsmath,amssymb,
 aps,
prb,
longbibliography,
 lengthcheck,%
]{revtex4-1}

\usepackage{graphicx}
\usepackage{dcolumn}
\usepackage{bm}
\usepackage{hyperref}
\usepackage{sidecap}
\usepackage[mathlines]{lineno}
\makeatletter
\newcommand*{\rom}[1]{\expandafter\@slowromancap\romannumerial#1@}
\makeatother

\begin{document}

\preprint{APS/123-QED}

\title{Infrared phonon anomaly and magnetic excitations in single-crystal Cu$_{3}$Bi(SeO$_{3}$)$_{2}$O$_{2}$Cl}
\author{K. H. Miller}
\affiliation{Department of Physics, University of Florida, Gainesville, Florida 32611-8440, USA}
\author{P. W. Stephens}
\affiliation{Department of Physics and Astronomy, Stony Brook University, Stony Brook, New York 11794-3800, USA}
\affiliation{Photon Sciences, Brookhaven National Laboratory, Upton, New York 11973, USA}
\author{C. Martin}
\affiliation{Department of Physics, University of Florida, Gainesville, Florida 32611-8440, USA}
\author{E. Constable}
\affiliation{Institute for Superconducting and Electronic Materials, University of Wollongong, New South Wales 2522, Australia}
\author{R. A. Lewis}
\affiliation{Institute for Superconducting and Electronic Materials, University of Wollongong, New South Wales 2522, Australia}
\author{H. Berger}
\affiliation{Institute of Physics of Complex Matter, Ecole Polytechnique Federal de Lausanne, CH-1015 Lausanne, Switzerland}
\author{G. L. Carr}
\affiliation{Photon Sciences, Brookhaven National Laboratory, Upton, New York 11973, USA}
\author{D. B. Tanner}
\affiliation{Department of Physics, University of Florida, Gainesville, Florida 32611-8440, USA}

\begin{abstract}
Infrared reflection and transmission as a function of temperature have been measured on single crystals of Cu$_{3}$Bi(SeO$_{3}$)$_{2}$O$_{2}$Cl.  The complex dielectric function and optical properties along all three principal axes of the orthorhombic cell were obtained via Kramers-Kronig analysis and by fits to a Drude-Lorentz model.  Below 115 K, 16 additional modes (8(E$\parallel\hat{a}$)+6(E$\parallel\hat{b}$)+2(E$\parallel\hat{c}$)) appear in the phonon spectra; however, powder x-ray diffraction measurements do not detect a new structure at 85 K.  Potential explanations for the new phonon modes are discussed.  Transmission in the far infrared as a function of temperature has revealed magnetic excitations originating below the magnetic ordering temperature ($T_{c}\sim$24 K).  The origin of the excitations in the magnetically ordered state will be discussed in terms of their response to different polarizations of incident light, behavior in externally-applied magnetic fields, and the anisotropic magnetic properties of Cu$_{3}$Bi(SeO$_{3}$)$_{2}$O$_{2}$Cl as determined by d.c.~susceptibility measurements.  

\end{abstract}

\maketitle

\section{Introduction}
Geometrically frustrated materials possessing magnetic order have garnered interest from both scientific and technological viewpoints due to their multiplicity of low energy nearly-equivalent states, which in turn prevent simple long-range magnetic order (e.g., antiferromagnetism).  More specifically, because neighboring antiferromagnetically-ordered moments in a frustrated structure cannot simultaneously minimize their local exchange energies, they are forced to cant with respect to their preferred directions.  The canting of magnetic moments in a frustrated system is known to produce noncollinear short-range magnetic orders\cite{NatMater.6.13} (e.g., cycloidal or spiral spin orders) that in turn yield many device-applicable phenomena such as magnetically driven ferroelectricity and large magnetoelectric responses.\cite{PRB.82.060402,PRL.92.257201}   In addition, the ability to synthesize geometrically frustrated magnetic materials in a layered structure facilitates the study of the intriguing phenomena associated with frustration because it allows for a quasi two-dimensional model of a three-dimensional system.  

Cu$_{3}$Bi(SeO$_{3}$)$_{2}$O$_{2}$Cl is a geometrically-frustrated layered material possessing magnetic order.  P. Millet {\it et al.\/}\cite{JMC.11.1152} determined the room temperature crystal structure of Cu$_{3}$Bi(SeO$_{3}$)$_{2}$O$_{2}$Cl using single crystal x-ray diffraction.  The Cu$_{3}$Bi(SeO$_{3}$)$_{2}$O$_{2}$Cl compound crystallizes in a layered structure where the individual layers stack along the $c$ axis of the orthorhombic cell ({\it Pmmn\/}).  The magnetic Cu$^{2+}$ ions form a hexagonal arrangement within each plane that is reminiscent of a Kagome lattice, and thus implies magnetic frustration.  Within any one hexagon there exist two unique copper sites, Cu1 and Cu2, which possess different out-of-plane oxygen bonding.  Both copper sites form distinct [CuO$_{4}$] units that are linked by Se$^{4+}$ and Bi$^{3+}$ ions.  The Cl$^{-}$ atoms, which sandwich the planes formed by the copper hexagons, position themselves along the axis that defines the parallel stacking of the copper hexagons.  Pictures and more detailed descriptions of the crystal structure are found elsewhere.\cite{JMC.11.1152,arXiv:1203.2782v1}

P. Millet {\it et al.\/}\cite{JMC.11.1152} also reported magnetic susceptibility measurements on a powder sample of Cu$_{3}$Bi(SeO$_{3}$)$_{2}$O$_{2}$Cl.  Near 150 K they observed a change in slope of 1/$\chi(T)$ that resulted in two distinct and positive Weiss temperatures.  To reconcile this anomaly in 1/$\chi(T)$, they performed linear birefringence on a single crystal and deduced that a second-order structural transition was likely, but they were not able to determine the exact nature of the transition.  P. Millet {\it et al.\/} also reported ferromagnetic-like behavior below T$_{c}$ $\approx$ 24~K.

The technique of infrared spectroscopy lends itself well to the investigation of geometrically frustrated magnetic materials.  For example, the spin-driven Jahn-Teller effect in the Cr spinel compounds CdCr$_2$O$_4$ and ZnCr$_2$O$_4$ acts to lift the degenerate ground state arising from competing magnetic interactions via a coupling to the lattice degrees of freedom; the ensuing lattice distortion is unambiguously observed in the infrared as a splitting of infrared-active phonon modes.\cite{PRB.80.214417,PRL.94.137202}  Furthermore, the a.c.~electric (magnetic) field of the infrared light can interact with ordered moments and excite an electromagnon (magnon) excitation, which in turn can provide information about the symmetry and nature of the magnetic order.\cite{PR.129.1566,NatPhys.2.97}  

Here we present our infrared studies on single crystal Cu$_{3}$Bi(SeO$_{3}$)$_{2}$O$_{2}$Cl.  We observe 16 new modes in the phonon spectra originating below 115~K.  Strikingly, our subsequent powder x-ray diffraction measurements reveal the same 300~K structure existing at 85~K.  Preliminary Raman measurements suggest that a loss of inversion symmetry is likely.  In addition, we observe new excitations in the infrared arising in the magnetically ordered state (below 24~K) that show isotropic infrared polarization dependence but anisotropic external magnetic field dependence.  In light of the novel excitations observed in the magnetically ordered state, we performed d.c.~susceptibility measurements to examine the anisotropic magnetic properties of Cu$_{3}$Bi(SeO$_{3}$)$_{2}$O$_{2}$Cl.  The results of our magnetic susceptibility measurements are in agreement with a recent report\cite{arXiv:1203.2782v1} of the magnetic properties in the similar Cu$_{3}$Bi(SeO$_{3}$)$_{2}$O$_{2}$Br compound (T$_{c}$ = 27.4~K).  (This manuscript is accompanied by supplementary material including additional figures referred to in the text and the results of full Rietveld refinements.)

\section{Experimental Procedures}
Single crystals of Cu$_{3}$Bi(SeO$_{3}$)$_{2}$O$_{2}$Cl were grown by standard chemical vapor-phase method. Mixtures of analytical grade purity CuO, SeO$_{2}$ and BiOCl powder in molar ratio 3:2:1 were sealed in quartz tubes with electronic grade HCl as the transport gas for the crystal growth. The ampoules were then placed horizontally into a tubular two-zone furnace and heated at 50$^{\circ}$C/h to 450$^{\circ}$C. The optimum temperatures at the source and deposition zones for the growth of single crystals were found to be 480$^{\circ}$C and 400$^{\circ}$C, respectively. After four weeks, many tabular green Cu$_{3}$Bi(SeO$_{3}$)$_{2}$O$_{2}$Cl crystals with a maximum size of $15\times12\times1$~mm$^3$ were obtained, which were indentified as synthetic francisite on the basis of x-ray powder diffraction data.

Zero-field temperature-dependent (7--300~K) reflectance and transmittance measurements were collected on a $6.7\times4.8\times0.7$~mm$^3$ single crystal (crystal 1) using a Bruker 113v Fourier transform interferometer in conjunction with a 4.2 K silicon bolometer detector in the spectral range 25--700 cm$^{-1}$ and a nitrogen cooled MCT detector from 700--5,000 cm$^{-1}$.  Room temperature measurements from 5,000--33,000 cm$^{-1}$ were obtained with a Zeiss microscope photometer.  Magnetic field-dependent transmission measurements in the spectral range 15--100 cm$^{-1}$ were performed on a $7.5\times3.8\times0.2$~mm$^3$ single crystal (crystal 2) at beam line U4IR of the National Synchrotron Light Source, Brookhaven National Laboratory, utilizing a Bruker IFS 66-v/S spectrometer.  The crystal was placed in a 10~T Oxford superconducting magnet and the transmitted intensities were measured using a 1.8 K silicon bolometer detector.  Additional magnetic field-dependent transmission measurements were performed on a $5.1\times3.5\times0.3$~mm$^3$ single crystal (crystal 3) in the spectral range 10--45 cm$^{-1}$ using a modified Polytec FIR 25 spectrometer interfaced to a 7~T Oxford split-coil superconducting magnet at the University of Wollongong.  The transmitted intensities were subsequently converted to absorption coefficients using a modified Beer-Lambert law that accounted for reflection losses at the surfaces.  Linearly polarized light was used in all infrared measurements and oriented along the three principal dielectric axes of the system.  For an orthorhombic system, the three principal dielectric axes coincide with the three crystallographic axes.\cite{Born&Wolf}  In the following report, the symbol ``$E$'' always refers to the polarization of the incoming light, while ``$H$'' is the orientation of the external field (when applicable).  

Powder x-ray diffraction measurements were performed at beamline X16C of the National
Synchrotron Light Source, Brookhaven National Laboratory.
A small crystal was crushed in a mortar and pestle, mixed with a small amount of Si powder (NIST Standard Reference Material 640c) and roughly three times its volume of ground cork, and loaded in a thin-walled glass capillary of  1 mm nominal diameter.
The cork served to dilute the sample so that it could fill the cross-section of the x-ray beam without drastically absorbing it; measured transmission at the center of the capillary was 12\%.  X-rays of nominal wavelength 0.6057 \AA\  were selected by a channel-cut Si(111) monochromator before the sample; small drifts in the x-ray wavelength were corrected via the Si internal standard.
The diffracted beam was analyzed by a Ge(111) crystal and detected by a commercial NaI(Tl) detector.  
Data were typically collected in steps of diffraction angle $2\theta$ of $0.005^{\circ}$.  For lattice parameter measurements as a function of temperature, the sample was inside a Be heat shield in a closed cycle He refrigerator, which was rocked several degrees at each $2\theta$ step.
Data for crystallographic refinements were collected with the sample continuously spinning several revolutions for each point, in an Oxford Cryostream sample cooler.
Powder x-ray diffraction data were analyzed using Topas-Academic software.\cite{topas}

Magnetic measurements were performed in a commercial superconducting quantum interference device magnetometer on crystal 2.  
 
\section{Results and Analysis}
\subsection{Zero Field Reflectance and Transmittance Spectra}
\begin{SCfigure*}
\includegraphics[width=0.75\textwidth]{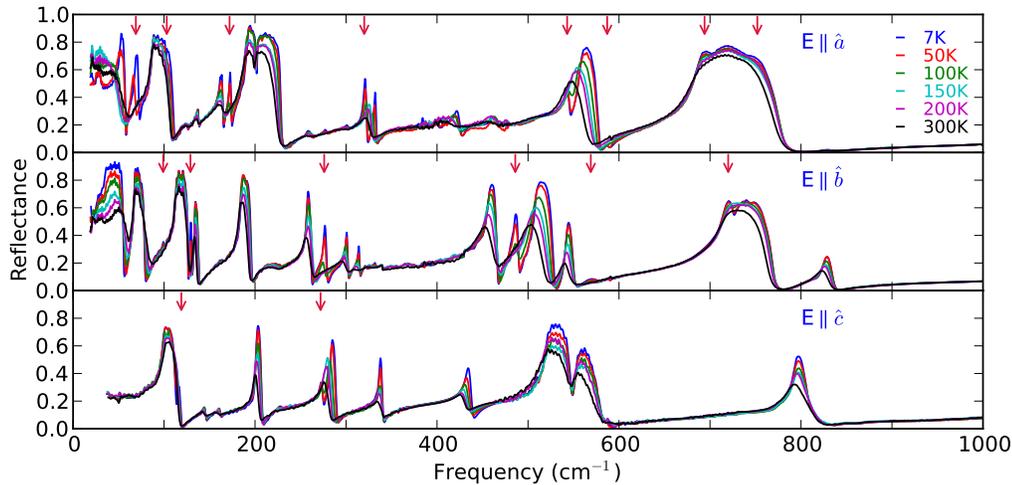}%
\caption{(Color online) The temperature-dependent reflectance spectra of Cu$_{3}$Bi(SeO$_{3}$)$_{2}$O$_{2}$Cl along the $a$, $b$, and $c$ axes.  Crimson arrows indicated the positions of new phonons arising at 115~K.}   
\label{Refl_temps}
\end{SCfigure*}

The temperature-dependent reflectance spectra of Cu$_{3}$Bi(SeO$_{3}$)$_{2}$O$_{2}$Cl along the $a$, $b$, and $c$ axes in the frequency interval 30--1,000~cm$^{-1}$ (4--120 meV) is shown in Fig.~\ref{Refl_temps}.  Remarkably, many new phonon modes are observed in the reflectance spectra along all three crystal directions upon cooling from 120~K to 110~K.  The arrows in Fig.~\ref{Refl_temps} indicate the position where new modes occur.  Meticulous temperature sweeps (every 1~K) between 120~K and 110~K indicate that all the new modes arise at 115 K; this trend was verified upon cooling and warming.  

In addition, a strong sharpening of many modes is observed with decreasing temperature.  In regions where new modes appear, the typical softening of resonance frequencies with increasing temperature is not systematically observed.  The typical softening of resonance frequencies with increasing temperature, which is correlated to the expansion of the lattice, can be counteracted by repulsion due to phonon mixing.  An explanation of the repulsion of phonons, which are bosonic excitations, will be presented in the discussion section.  In addition to the new phonons, we observe the softening (at 5~K to $\sim$30$\%$ of the 300 K value) of the lowest frequency mode along the $\hat b$ direction, an occurrence which is not associated with phonon repulsion.  The soft mode behavior exhibited by this low frequency phonon is reminiscent of that occurring in a displacive ferroelectric material.  

\begin{figure}[htp]
\includegraphics[width=0.45\textwidth]{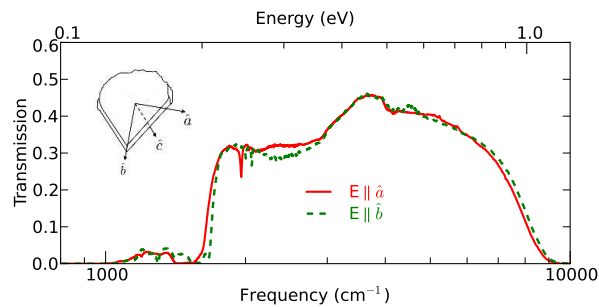}%
 \caption{(Color online) Transmission along the $a$ and $b$ axes of crystal 1 in the mid-infrared and near-infrared regions.  The inset is a sketch of crystal 1 with the $a$, $b$, and $c$ axes indicated.}   
\label{highTrans}
\end{figure}

Cu$_{3}$Bi(SeO$_{3}$)$_{2}$O$_{2}$Cl is opaque between $\sim$40 and 800~cm$^{-1}$ due to the strong optical phonon absorptions in this region; however, above the optical absorptions light is observed to transmit.  Figure~\ref{highTrans} displays the zero field 300~K transmission spectra of Cu$_{3}$Bi(SeO$_{3}$)$_{2}$O$_{2}$Cl along the $a$ and $b$ axes in the mid-infrared and near-infrared regions.  (Due to the geometry of the sample no light was observed to transmit along the $c$ axis.)  The anisotropy, which is prominent in the infrared reflectance spectra, is also observed in transmission at high frequencies.  Sharp but weak absorption features exist at 1950~cm$^{-1}$ along the $a$ axis and at 2000 and 2050~cm$^{-1}$ along the $b$ axis.  The aforementioned absorptions exhibit a minimal strengthening with decreasing temperature.  The downturn in transmission at 9,000 cm$^{-1}$ ($\sim$1.1 eV) indicates the onset of electronic absorptions.  The energy of the observed gap is consistent with the insulating nature of the material.  Transmission below the optical absorptions (below 40 cm$^{-1}$) is the subject of section~III~E.  

\subsection{Kramers-Kronig and Oscillator-model fits}
Along the $a$ and $b$ axes we have utilized a combined reflection and transmission analysis to extract the single bounce reflectance necessary for the Kramers-Kronig transformation.  More specifically, in the frequency region where the crystal transmits along the $a$ and $b$ axes, the measured reflectance has an additional contribution from the back surface.  Using the method employed by Zibold {\it et al.\/},\cite{PRB.53.11734} i.e., assuming that {\it k\/} is small in the region of interest, we are able to extract the single bounce reflectance of the sample.  Since the $c$ axis does not exhibit transmission, the measured reflectance was assumed to be single bounce reflectance along this direction.  

The real and imaginary parts of the dielectric function were estimated from the single bounce reflectance, $R_{s}(\omega$), using the Kramers-Kronig transformation.\cite{Wooten}  Before calculating the Kramers-Kronig integral, the low frequency (0.1--30 cm$^{-1}$) data were approximated using the dielectric function determined from the fitting procedure described below.  At high frequencies the reflectance was assumed to be constant up to $1 \times 10^7$~cm$^{-1}$, after which $R\sim(\omega)^{-4}$ was assumed as the appropriate behavior for free carriers. The optical properties were obtained from the measured reflectance and the Kramers-Kronig derived phase shift on reflection. 

The single bounce reflectance was fit with a Lorentz oscillator model to obtain a second estimate of the complex dielectric function in the infrared range.  The model assigns a Lorentzian oscillator to each phonon mode in the spectrum plus a high frequency permittivity, $\epsilon_\infty$, to address the contribution of electronic absorptions.  The model has the following mathematical form:
\begin{equation}
\epsilon(\omega) = \displaystyle\sum_{j=1}^{\infty}\frac{S_j\omega_j^2}{{{\omega_j}^2}-{{\omega}^2}-{i\omega\gamma_j}} +\epsilon_\infty\;\;\;,
\label{epsilonDL}
\end{equation}
where $S_j$, $\omega_j$, and $\gamma_j$ signify the oscillator strength, center frequency, and the full width at half max (FWHM) of the j$^{\text{th}}$ Lorentzian oscillator.  The reflectivity is then calculated using the Drude-Lorentz complex dielectric function.  A figure comparing the calculated and measured reflectivity is available in the supplementary information.

\subsection{Optical Properties}
\begin{figure}[h]
\includegraphics[width=0.50\textwidth]{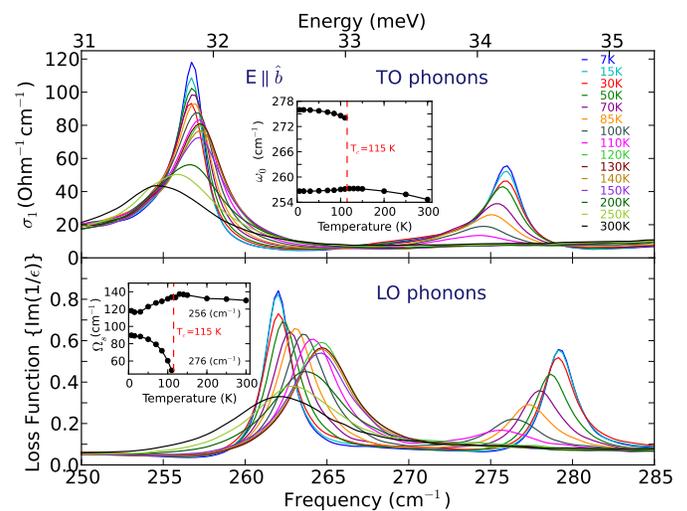}%
 \caption{(Color online) The temperature dependent optical conductivity $\sigma_{1}(\omega)$ (upper panel) and loss function Im(-1/$\epsilon(\omega)$) (lower panel) along the $b$ axis in the frequency range 250--285~cm$^{-1}$.  The insets of the upper and lower panels depict respectively the temperature dependence of the resonance and plasma frequencies of the two modes.}  
\label{cond_lossfx}
\end{figure}

The optical properties extracted from both Kramers-Kronig analysis and our fits of reflectance can be used to investigate further the appearance of new phonon modes in the infrared spectra at 115 K.  The upper panel of Fig.~\ref{cond_lossfx} depicts the real part of the optical conductivity, $\sigma_{1}(\omega)$, along the $b$ axis in the frequency range 250--285 cm$^{-1}$.  The lower panel of Fig.~\ref{cond_lossfx} depicts the loss function, Im(-1/$\epsilon(\omega)$), in the same region.  Both optical properties shown are from Kramers-Kronig analysis of the single bounce reflectance.  The peaks observed in $\sigma_{1}(\omega)$ and Im(-1/$\epsilon(\omega)$) closely correspond respectively to the TO and LO phonon frequencies.  The inset of the lower panel of Fig.~\ref{cond_lossfx} depicts the plasma frequency ($\Omega_s=\sqrt{S_{j}\omega_{0}^{2}}$) associated with both phonons as a function of temperature as obtained by our Drude-Lorentz fitting.  The equation defining the plasma frequency of ionic vibrations is similar in form to that which describes the free carrier response in metallic systems after the electronic mass and charge are replaced with the reduced mass of the normal mode and ionic charge.  The plasma frequency is of particular interest here because its square is proportional to the spectral weight associated with a particular phonon.  The new phonon mode that appears at $\sim$ 276~cm$^{-1}$ below 115~K gains spectral weight at the expense of the existing mode at $\sim$ 256~cm$^{-1}$.  The smooth shift of spectral weight with decreasing temperature from the existing mode to the new mode is reminiscent of a second order transition.

\subsection{Powder X-Ray Diffraction}
\begin{figure}[h]
\includegraphics[width=0.50\textwidth]{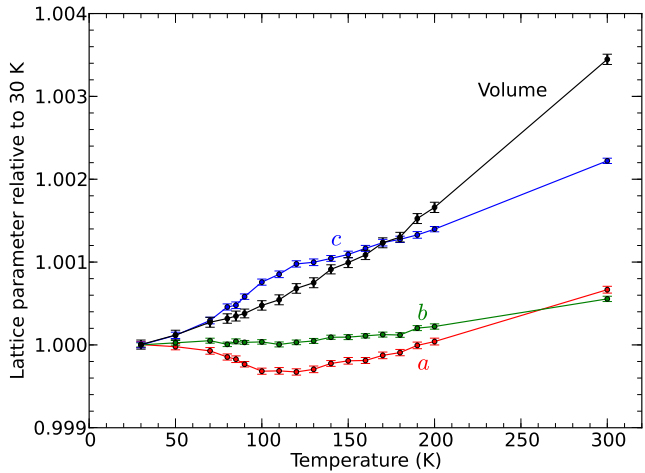}%
 \caption{(Color online) Lattice parameters and unit cell volume \textit{vs.} temperature.  
Error bars reflect the reproducibility of independently measured data sets at selected temperatures.
Solid lines are drawn as guides to the eye.  Lattice parameters at 30 K are $a$ = 6.3463(2) \AA, $b$ = 9.6277(3) \AA, $c$ = 7.2186(3) \AA, Volume = 441.10(2) \AA$^3$.}  
\label{lattice_parms}
\end{figure}

The changes in the infrared spectra, namely the appearance of new vibrational modes below 115~K, suggest a reduction in lattice symmetry.
To investigate the low temperature structure of Cu$_{3}$Bi(SeO$_{3}$)$_{2}$O$_{2}$Cl, we performed powder x-ray diffraction measurements between 30 and 300~K.  
The lattice parameters as a function of temperature extracted from Rietveld fits of the diffraction spectra are shown in Fig.~\ref{lattice_parms}.
There exists some structure in the curves of lattice parameters \textit{vs.} temperature, but there is nothing that could be regarded as conclusive evidence of a phase transition.
The negative thermal expansion observed for the $a$ lattice parameter below 100~K is not unusual for an ionic compound of orthorhombic symmetry where it may be energetically favorable for the unit cell to contract in one direction while expanding in other directions.

A figure depicting Rietveld refinements of the powder x-ray diffraction pattern at 295 and 85~K is available in the supplementary information.  Both refinements were consistent with the published structure of P. Millet \textit{et al.} in space group $Pmmn$.  
Full details of the refinements, including bonding geometry, may also be found in the supplementary information.
We saw no evidence for splitting or broadening of diffraction lines nor the appearance of new diffraction lines at low temperature.  
We conclude that there is no direct evidence for a lowering of crystallographic symmetry near 115~K.
This is a serious puzzle because the appearance of 16 new infrared modes below 115~K implies that the symmetry of the crystal is lower than $Pmmn$ below that temperature.
\subsection{Magnetic Field-Dependent Transmission}
Far-infrared transmission as a function of temperature and external magnetic field was measured on crystal 2 and crystal 3 with light polarized along the $a$ axis, $b$ axis, and at $45^{\circ}$ to both the $a$ and $b$ axes.  Upon cooling to 5~K an excitation was observed at 33.1~cm$^{-1}$ that only existed in  the magnetically ordered state (below 24 K).  The excitation was observed in all four polarizations.  A figure depicting the excitation, at 5~K, in each of the four polarizations measured, as well as the evolution of the excitation as the external field is ramped to 10~T ($H\parallel\hat{c}$ geometry) is available in the supplementary information.

\begin{figure}[h]
\includegraphics[width=0.50\textwidth]{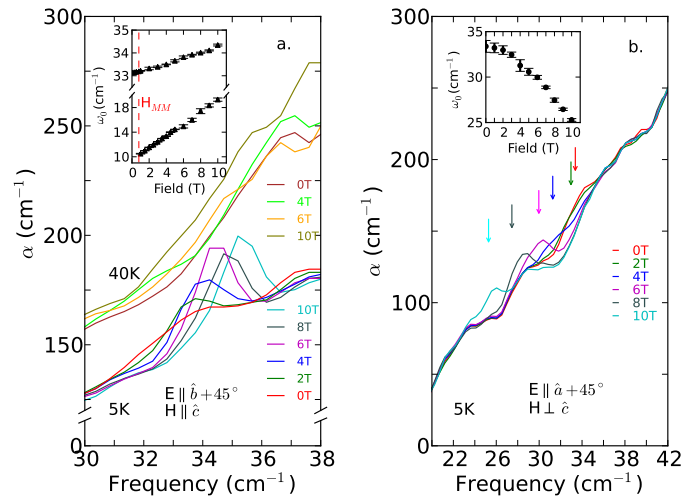}%
 \caption{(Color online) Experimental support (a) for claiming that the magnetic resonance observed at 33.1 cm$^{-1}$ exists below the magnetic ordering temperature (T$_{c}\sim$24~K).  External fields, which act to sharpen the resonance, were also applied above T$_{c}$ for further verification that the excitation was not observed.  In the inset, the dashed red line denoted H$_{MM}$ indicates the region where the metamagnetic transition occurs in the H$\parallel\hat{c}$ geometry.  The isotropic magnetic excitation's field dependence in the $H\perp\hat{c}$ geometry (b) with light polarized along the E$\parallel\hat{a}+45^{\circ}$ direction.}  
\label{proof_HperpC}
\end{figure}

When the external magnetic field was applied parallel to the $c$ axis ($H\parallel\hat{c}$) and the field was ramped to 10~T, the excitation at 33.1~cm$^{-1}$ increased linearly with increasing field (inset Fig.~\ref{proof_HperpC}a).  No hysteresis effects were observed upon ramping the magnetic field.  To establish confidence that the observed excitation only existed in the magnetically ordered state, the crystal was warmed up to 40 K and the external field was ramped from 0 to 10 T.  The excitation was not observed.  The results are shown in Fig.~\ref{proof_HperpC}a for the E$\parallel\hat{b}+45^{\circ}$ polarization.

\begin{figure}[h]
\includegraphics[width=0.50\textwidth]{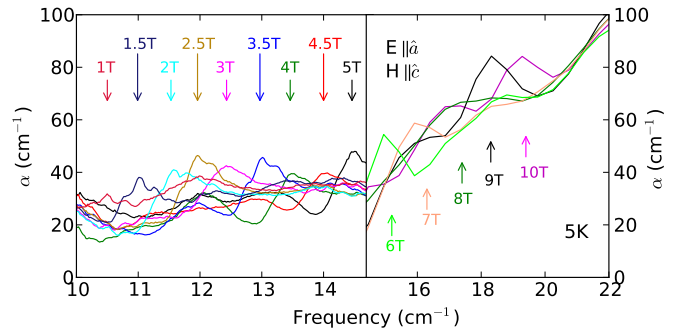}%
 \caption{(Color online) The lower frequency magnetic resonance mode observed in the H$\parallel\hat{c}$ geometry for fields of 1~T and greater.  The two panels result from complimentary techniques on two separate crystals of Cu$_{3}$Bi(SeO$_{3}$)$_{2}$O$_{2}$Cl.  Experimental limitations prevented a study of the mode below 1~T fields.}  
\label{lowmode}
\end{figure}

In addition, for fields 1~T and greater applied parallel to the $c$ axis, a second magnetic excitation was observed which also possessed isotropic polarization dependence in the $ab$ plane.  The excitation increased linearly from 10.5~cm$^{-1}$ at 1~T to 19.3~cm$^{-1}$ at 10~T.  Because of the low signal to noise level it is not known whether the excitation disappeared below 1~T or it was just unresolved.  The excitation is pictured for the $E\parallel\hat{a}$ polarization in Fig.~\ref{lowmode}.  The spectra in Fig.~\ref{lowmode} is a conjunction of data from two different interferometers on two separate crystals of Cu$_{3}$Bi(SeO$_{3}$)$_{2}$O$_{2}$Cl (see Experimental Procedures section).  The complimentary techniques on separate crystals establish confidence in the existence of the excitation.  

When the external magnetic field was applied perpendicular to the $c$ axis ($H\perp\hat{c}$) and ramped to 10 T, the excitation at 33.1~cm$^{-1}$ decreased its resonance frequency quadratically with field (inset Fig.~\ref{proof_HperpC}b).  The field dependence of the excitation for the E$\parallel\hat{a}+45^{\circ}$ polarization is depicted in Fig.~\ref{proof_HperpC}b.  It should be noted that the details of our experimental setup required that the external magnetic field be applied parallel to the polarization of the incoming light in the $H\perp\hat{c}$ geometry.  No additional modes in this orientation of external field were detected.

Similar experiments in magnetic fields were carried out in reflectance geometry; however, the excitations were much too weak to give rise to reflectance bands. 
\subsection{Magnetic Properties}
\begin{figure}[h]
\includegraphics[width=0.50\textwidth]{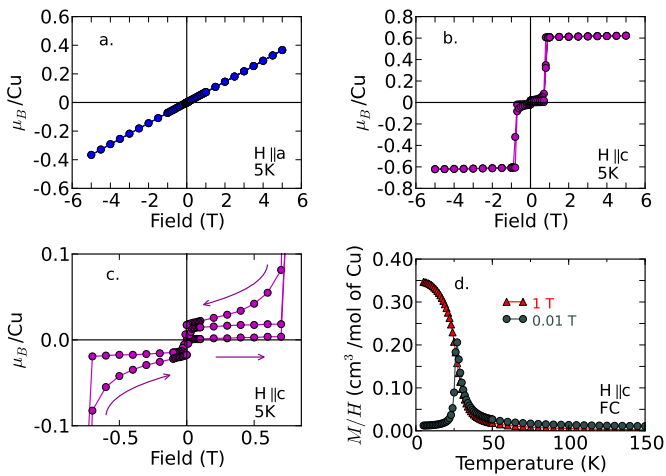}%
 \caption{(Color online) Isothermal magnetization measurements at 5 K, which is well within the magnetically ordered state, for $H\perp\hat{c}$ (a) and $H\parallel\hat{c}$ (b,c).  The magnetization as a function of temperature (d) for $H\parallel\hat{c}$ at fields above and below the metamagnetic transition as measured while warming after field cooling to 5~K.}  
\label{MH}
\end{figure}

The anisotropic response to external magnetic fields of the magnetic excitation  observed at 33.1~cm$^{-1}$ in transmission has inspired an investigation of the anisotropic magnetic properties of Cu$_{3}$Bi(SeO$_{3}$)$_{2}$O$_{2}$Cl through d.c.~susceptibility measurements.  The results of our d.c.~susceptibility measurements are depicted in Fig.~\ref{MH}.  Isothermal magnetization measurements taken at 5 K exhibit the strong anisotropic response to the direction of an external field previously noted in transmission measurements.  As shown in Fig.~\ref{MH}a, with $H\perp\hat{c}$, the resulting magnetization \textit{vs.} field loop resembles that of an antiferromagnet.  On the contrary, with $H\parallel\hat{c}$ (Fig.~\ref{MH}b), antiferromagnetic behavior is observed from 0 to 0.1~T, followed by a metamagnetic transition from 0.1 to 0.8~T, and then ferro or ferrimagnetic behavior persists from 0.8 to 5~T.  Figure~\ref{MH}c is an enlarged view of the metamagnetic transition in the $H\parallel\hat{c}$ geometry.  The hysteresis observed upon sweeping the field is likely associated with the metamagnetic transition.  To verify further the antiferromagnetic to ferro or ferrimagnetic transition occurring in the $H\parallel\hat{c}$ geometry, magnetization was measured as a function of temperature in fields of 0.01 and 1 T.  The results are shown in Fig.~\ref{MH}d.  The low temperature cancellation of oppositely aligned moments expected for an antiferromagnet is observed at 0.01 T, whereas a strong ferro or ferrimagnetic magnetization is observed at low temperature for the 1~T magnetization data.  

Our results and interpretation are in qualitative agreement with a recent report by M. Pregelj {\it et al.\/}\cite{arXiv:1203.2782v1} on the similar Cu$_{3}$Bi(SeO$_{3}$)$_{2}$O$_{2}$Br compound ($T_{c}$ = 27.4~K).  Small differences in the magnetization saturization value with $H\parallel\hat {c}$ is likely linked to defects arising from different growth conditions.  We normalize our susceptibility curves to moles of Cu while M. Pregelj {\it et al.\/} normalize to moles of formula unit (scaling factor of three difference).

\begin{figure}[h]
\includegraphics[width=0.50\textwidth]{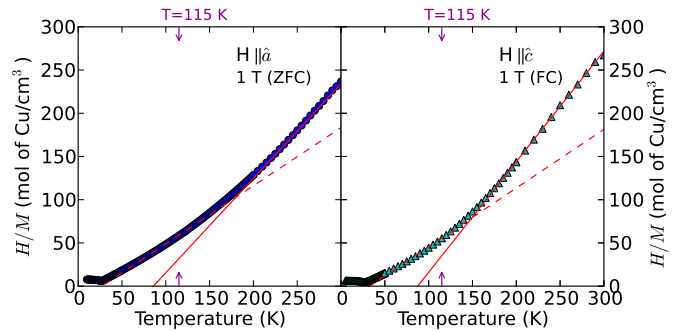}%
 \caption{(Color online) Plot of $H/M$ as a function of temperature measured with a 1 T magnetic field oriented parallel to the $a$ axis (left panel), and also parallel to the $c$ axis (right panel).}  
\label{CHIINV}
\end{figure}

The origin of the anomaly in 1/$\chi(T)$, first observed in powder samples by P. Millet {\it et al.\/},\cite{JMC.11.1152} and subsequently observed by us along multiple axes of single crystals of Cu$_{3}$Bi(SeO$_{3}$)$_{2}$O$_{2}$Cl (Fig.~\ref{CHIINV}) and Cu$_{3}$Bi(SeO$_{3}$)$_{2}$O$_{2}$Br,\cite{arXiv:1203.2782v1} remains an unresolved issue.  Below 150~K, the plot of $H/M$ exhibits a slight curvature.  We therefore conclude that the data is no longer accurately described by the Curie-Weiss law below 150~K (a dashed line fit below 150~K remains in Fig.~\ref{CHIINV} to exemplify the discrepancy).  Such deviations from Curie-Weiss law can be expected for quasi-two-dimensional spin systems and for frustrated topologies where non-trivial spin-correlation functions characterize the ``classical spin liquid regime'' or ``correlated paramagnet'' between the Curie-Weiss temperature (from the Curie-Weiss fit above 150~K) and the magnetic ordering temperature (24~K).

We observed the anomaly for external fields of 1~T and 0.01~T in both zero field cooling as well as field cooling measurements.

 \section{Discussion}
\subsection{Group Theory and Observed Modes}
The number of optical modes in the Cu$_{3}$Bi(SeO$_{3}$)$_{2}$O$_{2}$Cl compound can be determined by group theory analysis.  Using the SMODES\cite{SMODES} program, we arrive at the following distribution of modes: 
\begin{eqnarray*}
\Gamma^{optical}=14B^{\it(IR)}_{1u}+14B_{2u}^{\it(IR)}+11B_{3u}^{\it(IR)}\\
+12A_{g}^{\it(R)}+6B_{1g}^{\it(R)}+9B_{2g}^{\it(R)}+12B_{3g}^{\it(R)}+9A_{1u}
\label{Gamma_optical}
\end{eqnarray*}
where (R) and (IR) denote respectively Raman active and infrared active modes.  The B$_{3u}$, B$_{2u}$, and B$_{1u}$ modes are infrared active along the ${a}$, ${b}$, and ${c}$ crystal axes respectively.  The 9A$_{1u}$ modes are silent. The 12A$_{g}$ modes are Raman active radial breathing modes and require parallel incoming and outgoing polarizations according to the selection rules of the {\it Pmmn\/} spacegroup.  The B$_{3g}$, B$_{2g}$, and B$_{1g}$ modes are also Raman active, but require crossed polarization of the light ($bc$, $ac$, and $ab$ respectively).

At 300 K we observe all 11 of the 11 predicted B$_{3u}$ modes along the $a$ axis, all 14 of the 14 predicted B$_{2u}$ modes along the $b$ axis, and 11 of the 14 predicted B$_{1u}$ modes along the $c$ axis.  The discrepancy between observed and detected modes along the $c$ axis is likely an experimental shortcoming arising from the low signal to noise when measuring along the thin dimension of the crystal.  At 115~K we detect 8 additional modes along the $a$ axis, 6 additional modes along the $b$ axis, and 2 additional modes along the $c$ axis.  To study the observed modes, the reader is referred to Table \ref{Table} where the modes at 7~K along all three crystal axes are identified as well as their respective Lorentz oscillator parameters.  An asterisk after the TO frequency indicates a new mode that arises at 115~K.  

In materials with an inversion center, the \textbf{k} = 0 optical modes as measured through Raman and infrared reflectance are mutually exclusive.  When the inversion center in a structure is removed, the local centers of symmetry about which all the Raman modes have zero net dipole moment will be removed, and all Raman modes will therefore become infrared active.\cite{PRB.82.214302}  To further investigate the 16 new modes observed at 115~K, preliminary 300~K Raman measurements were recorded\cite{D.J.Arenas} in the $ab$ plane of crystal 1.  Raman modes observed at 172.9, 323.5, and 484.4~cm$^{-1}$ closely correspond to three of the new infrared modes observed below 115 K.  If we take into account the fact that Raman modes can slightly increase their resonance frequencies upon cooling to low temperatures, then 300 K Raman modes at 538.3 and 583.1~cm$^{-1}$ draw close redolence to two additional new modes observed in the infrared below 115 K.  We therefore turn to a closer examination of a possible centrosymmetric to non-centrosymmetric transition in Cu$_{3}$Bi(SeO$_{3}$)$_{2}$O$_{2}$Cl near 115 K.
\begin{table*}[htp]
\caption{Oscillator parameters for the infrared observed modes of Cu$_{3}$Bi(SeO$_{3}$)$_{2}$O$_{2}$Cl (at 7 K) along all three crystal axes.  The new modes arising below 115 K are indicated with an asterisk next to their corresponding TO frequencies.}
\begin{ruledtabular}
\begin{tabular}{l l l l || l l l l || l l l l}
 & $\hat {a}$ & && & $\hat {b}$ & &&  & $\hat {c}$ & \\
\hline
Osc Str & TO Freq & LO Freq & FWHM & Osc Str & TO Freq & LO Freq & FWHM & Osc Str & TO Freq & LO Freq & FWHM \\
S & $\omega$ (cm$^{-1}$) & $\omega$ (cm$^{-1}$) & $\gamma$ (cm$^{-1}$) & S & $\omega$ (cm$^{-1}$) & $\omega$ (cm$^{-1}$) & $\gamma$ (cm$^{-1}$) & S & $\omega$ (cm$^{-1}$) & $\omega$ (cm$^{-1}$) & $\gamma$ (cm$^{-1}$)\\
\hline
   3.947 & 52.8 & 57.7 & 1.7 &26.221 & 36.3 & 55.6 & 1.9 & 0.035 & 53.2 & 53.6 & 1.5\\
   1.291 & 69.9$\ast$&72.4 & 1.9 &2.389 & 68.3&78.8 & 1.3& 2.228 & 99.8&112.9 & 3.9\\
   4.096 & 89.0&101.9 & 1.9 &0.034 & 99.8$\ast$&100.2 & 1.0& 0.055 & 115.1$\ast$&118.4 & 2.8\\
   0.126 & 101.1$\ast$&110.7 & 5.7 &0.468 & 115.2&126.2 & 2.4& 0.059 & 144.8&145.8 & 3.0\\
   0.029 & 137.6&138.2 & 1.4 &0.071 & 128.9$\ast$&130.5 & 0.9& 0.023 & 161.5&162.0 & 1.9\\
   0.433 & 161.9&164.8 & 2.6 &0.154 & 133.5&138.2 & 1.3& 0.210 & 204.0&208.1 & 1.6\\
   0.223 & 172.3$\ast$&174.1 & 1.9 &0.744 & 185.8&196.2 & 1.2& 0.035 & 273.8$\ast$&274.7 & 1.5\\
   1.540 & 191.6&201.4 & 1.0 &0.211 & 256.9&261.9 & 1.2& 0.204 & 284.4&289.8 & 2.4\\
   0.115 & 202.1&228.3 & 3.1 &0.106 & 276.1$\ast$&279.2 & 2.0& 0.093 & 337.6&340.5 & 2.0\\
   0.085 & 320.0$\ast$&323.1 & 1.8 & 0.062 & 300.3&302.3 & 1.6& 0.154 & 433.5&439.3 & 4.7\\
   0.045 & 331.1&333.6 & 2.1 &0.038 & 313.9&315.2 & 1.6& 0.586 & 528.4&547.4 & 4.8\\
   0.067 & 422.9&426.0 & 4.7 &0.291 & 456.3&466.4 & 1.9& 0.145 & 554.4&579.4 & 12.4\\
   0.055 & 470.2&471.6 & 12.2 &0.156 & 484.7$\ast$&489.9 & 3.8& 0.136 & 794.9&812.7 & 6.3\\
   0.139 & 542.4$\ast$&546.1 & 5.0 &0.336 & 507.0&528.0 & 3.7&  &&  &  \\
   0.293 & 557.3&577.5 & 4.0 &0.059 & 542.3&550.2 & 3.1&  &&  & \\
   0.066 & 587.2$\ast$&587.4 & 3.9 &0.045 & 571.4$\ast$&575.7 & 22.0&  &&  &  \\
   0.564 & 688.2&669.1 & 7.1 &0.122 & 716.1$\ast$&725.9 & 5.3&  &&  &  \\
   0.122 & 703.8$\ast$&705.5 & 16.8 &0.062 & 730.0&767.6 & 10.3&  &&  &  \\
   0.005 & 737.0$\ast$&774.0 & 19.9 &0.024 & 811.3&815.0 & 16.1&  &&  &  \\
    &  &&  &0.030 & 825.0&834.4 & 7.9&  &&  &  \\
\end{tabular}
\end{ruledtabular}
\label{Table}
\end{table*} 

\subsection{Powder X-Ray diffraction}
An outstanding problem in the interpretation of our results is the fact that the powder x-ray diffraction measurements did not find any evidence for a phase transition accompanying the dramatic appearance of 16 infrared-active modes below 115~K.  In their 2001 report, P. Millet {\it et al.\/}\cite{JMC.11.1152} sought to detect a potential structural phase transition by optical birefringence, and they placed an upper limit of 1$^{\circ}$ on the possible rotation of optical axis, thus indicating no distortion to monoclinic symmetry at that level.  The present powder diffraction measurements would be sensitive to a monoclinic distortion on the order of 0.001$^{\circ}$, and none is observed.  Their analysis of linear birefringence suggest that the nature of the suspected transition is second-order, which is consistent with what we have found in our analysis of shifts in spectral weight from existing modes to new modes (inset lower panel Fig.\ref{cond_lossfx}).  In what follows, we consider lower-symmetry non-centrosymmetric orthorhombic structures that might be a potential host for Cu$_{3}$Bi(SeO$_{3}$)$_{2}$O$_{2}$Cl.  In addition, we scrutinize the possibility of a transition to an incommensurate lattice, and we discuss the implications of the recent neutron diffraction report on Cu$_{3}$Bi(SeO$_{3}$)$_{2}$O$_{2}$Br.\cite{arXiv:1203.2782v1}

The centrosymmetric $Pmmn$ space group has extinction class $P$-\,-$n$, i.e., it obeys the condition that $h+k$ must be even for $(hk0)$ reflections.  A continuous transition to a non-centrosymmetric orthorhombic subgroup of the same lattice dimensions could lead to space groups $P2_12_12$, $Pmm2$, $Pm2_1n$, or $P2_1mn$.  The two former choices would allow additional powder x-ray diffraction peaks.  We have carefully searched for them throughout the temperature range below 115~K without success.  The two latter space groups have the same extinction class as $Pmmn$, and so they represent a mechanism for breaking of inversion symmetry without producing a qualitative change in the powder diffraction pattern.  Such a distortion might not be easy to recognize from powder diffraction data because the atoms would presumably move a small distance from their undistorted locations, and so the $Pmmn$ model would still give a reasonably accurate description of the data in the acentric phase.
One possibility is to look for unusual behavior in thermal displacement parameters, similar to the method used to decode complicated distortions in magnetically frustrated spinels.\cite{Radaelli1}  We have not been able to detect such an effect from the data at hand.

Neutron powder diffraction measurements on Cu$_{3}$Bi(SeO$_{3}$)$_{2}$O$_{2}$Br at 60~K were refined in $Pmmn$,\cite{JMC.11.1152,arXiv:1203.2782v1} but that has only limited bearing on the present issue.  
First, it is not known that data from the bromide material suggests the same loss of inversion symmetry that we have reported here for the chloride material.
Second, as noted above, the difference between centric and acentric structures could be very difficult to see in neutron or x-ray powder diffraction.
Indeed, it is worth pointing out that the magnetic transition studied in the bromide analog may well have occurred from a symmetry lower than $Pmmn$.\cite{JMC.11.1152,arXiv:1203.2782v1}

Another possibility is that an incommensurate transition occurs in Cu$_{3}$Bi(SeO$_{3}$)$_{2}$O$_{2}$Cl below 115~K.  Incommensurate phases acquire satellite x-ray diffraction lines around the Bragg peaks of the symmetric phase.\cite{PR.185.211}  We did not observe satellite peaks, but it is possible they exist and were too weak to be detected in the present powder x-ray diffraction measurements.  We plan to perform more sensitive measurements (\text{e.g.,} single crystal or neutron powder diffraction) to resolve this issue.  

\subsection{Phonon Repulsion}
\begin{figure}[h]
\includegraphics[width=0.50\textwidth]{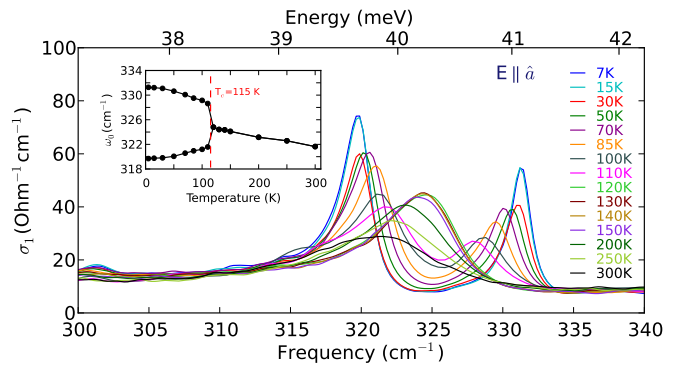}%
 \caption{(Color online) The real part of the optical conductivity along the $a$ axis in the frequency range 300--340~cm$^{-1}$.  At 115~K a new mode appears around $\sim$328 cm$^{-1}$ that is in close proximity to an existing mode at $\sim$323 cm$^{-1}$.  The repulsion of the two modes as temperature is decreased is emphasized in the inset where the Drude-Lorentz resonance frequencies are plotted as a function of temperature.}  
\label{repulsion}
\end{figure}

The typical dispersion of phonon resonance frequencies (i.e., softening with increasing temperature) as dictated by the anharmonic term in the lattice potential is not observed for a number of the new modes and existing modes in close proximity to the new modes appearing below 115~K.  According to our presumption that Raman modes become infrared active with a loss of inversion symmetry, we examined the physics behind phonons of similar strengths resonating at contiguous frequencies.  Figure~\ref{repulsion} depicts a new mode appearing around $\sim$328~cm$^{-1}$ along the $a$ axis that is in close proximity to an existing mode at $\sim$323~cm$^{-1}$.  As further emphasized in the inset of Fig.~\ref{repulsion}, the two modes strongly repel one another as temperature decreases, which seems uncharacteristic of bosonic excitations.  The repulsion is due to phonon mixing and can be understood by an analogy to the classical system of two coupled ideal harmonic oscillators.  The classical problem amounts to solving for the eigenvalues of a matrix with off diagonal terms arising from the coupling between the oscillators.  If both oscillators are given the same initial frequency (analogous to both phonons resonating at the same energy) and the coupling is turned on, then the oscillators will repel one another.  The repulsion increases with increasing coupling.  Applying the same concept to the case of phonon mixing, one can see that the coupling between the phonons is increased with decreasing temperature, an effect which is well understood via the thermal contraction of the lattice.  It is also worthwhile noting that the new phonon mode depicted in Fig.~\ref{cond_lossfx} at $\sim$276~cm$^{-1}$ causes a repulsion of the resonant frequency of the existing mode at $\sim$256~cm$^{-1}$ below 115~K.  The two modes appear to be initially non-degenerate because of their separation in energy ($\sim$20~cm$^{-1}$); however, as seen in the inset of Fig.~\ref{cond_lossfx} lower panel, there is certainly an interaction between the two modes.  The coupling is verified further by the shifts in oscillator strengths between the two modes (lower panel Fig.~\ref{cond_lossfx} inset).

\subsection{Magnetic Excitations}
Magnetic excitations stimulated by infrared light can be grouped into one of two categories:~(a) traditional magnons that are excited by the a.c.~magnetic field of the light, and (b) novel electromagnons that are excited by the a.c.~electric field component of the light.  Single-magnon absorptions, which at infrared frequencies are commonly observed in antiferromagnets and thusly tabbed antiferromagnetic resonance modes (AFMR), are magnetic dipole transitions that occur when the oscillating frequency of the light corresponds to the \textbf{k} = 0 frequency of a spin wave.  Electromagnons were first proposed as strongly renormalized spin waves with dipolar momentum.\cite{NatPhys.2.97}  Since their discovery, electromagnons have been extensively studied in rare earth manganite compounds, namely in TbMnO$_{3}$, where a large body of recent literature (Raman\cite{PRB.81.054428}, neutron\cite{PRL.98.137206}, and infrared\cite{PRL.101.187201}) has tied the excitation to the lattice itself.  The work has resulted in a generalized hybrid magnon-phonon mode picture of electromagnons.  Challenges arise in differentiating the two aforementioned excitations because they resonate in the same general frequency intervals.  A common solution is to measure the different faces of a crystal while rotating the polarization of the incoming light.  

The geometry of the Cu$_{3}$Bi(SeO$_{3}$)$_{2}$O$_{2}$Cl crystal only allowed for transmission measurements with the \textbf{k} vector of the incoming light aligned perpendicular to the {\it ab\/}  plane.  However, a thorough polarization study within the {\it ab\/} plane was carried out and it has lead to the observation of an isotropic magnetic excitation at 33.1~cm$^{-1}$.  These results contradict the previously reported anisotropic nature of magnons and electromagnons.  Moreover, when external magnetic fields are applied in the $H\parallel\hat{c}$  and $H\perp\hat{c}$ geometries, the 33.1~cm$^{-1}$ resonance shifts to higher and to lower frequencies respectively.  In what follows we will examine the nature of the excitation as well as propose a reason for its isotropic behavior.  External field dependent spectra with $H\parallel\hat{c}$ and $H\perp\hat{c}$ will be discussed separately.  (The following section will focus on the 33.1~cm$^{-1}$ excitation because it is the only mode observed in zero field.)

\subsubsection{Nature and Isotropy}

To elucidate the nature of the magnetic excitation observed in Cu$_{3}$Bi(SeO$_{3}$)$_{2}$O$_{2}$Cl at 33.1~cm$^{-1}$, we compare the strength of the observed excitation to the extensive literature on magnons and electromagnons in the infrared.  Our magnetic excitation creates a peak in $\alpha(\omega)$ that is $\sim$ 30~cm$^{-1}$ above the baseline.  Taking the average d.c.~index of refraction to be 3 in the $ab$ plane, we determine that our magnetic excitation corresponds to an optical conductivity of about 0.23 $\Omega^{-1}$cm$^{-1}$ ($\sigma_1=\frac{c}{4\pi}n\alpha$), which is roughly the same strength as other reported single-magnon excitations.\cite{PR.138.A1769,JAP.38.1496,PRB.47.5300}  Electromagnons, which have been extensively studied in a number of rare-earth manganites, are observed to possess optical conductivities of at least a factor of 10 larger.\cite{PRB.78.104414,NatPhys.2.97}  Although this method of comparison does not yield objective certainty, we can further support its conclusion by employing optical sum rule analysis on our measured reflectance data.  Since an electromagnon contributes to the dielectric constant, it must gain spectral weight from a dipole active excitation, the main candidates being domain relaxations, phonons, or electronic transitions.  The reflectivity spectra measured at 7 K and 30 K do not show any change associated with the magnetic excitation.  Therefore, the magnetic excitation does not gain spectral weight from the low frequency infrared-active phonon modes.   
\begin{figure}[h]
\includegraphics[width=0.50\textwidth]{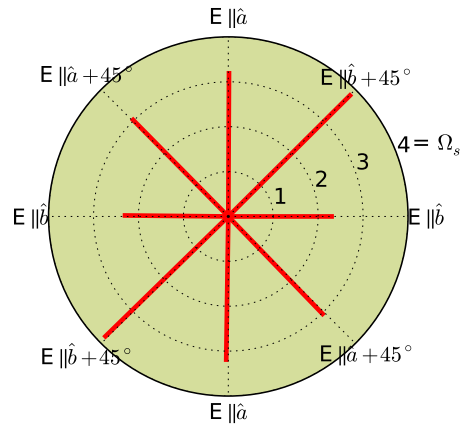}%
 \caption{(Color online) A polar plot of the oscillator strengths associated with the 33.1~cm$^{-1}$ mode at 0~T for all four polarizations measured in the $ab$ plane.}  
\label{polar0T}
\end{figure}

The isotropic character of the magnetic excitation observed at 33.1~cm$^{-1}$ is depicted in Fig.~\ref{polar0T} where the oscillator strength of the excitation is plotted versus the polarization angle of the light in the {\it ab\/} plane.  (It is worthwhile noting that infrared reflectance spectra measured on crystal 2 revealed the same strong anisotropy as depicted in Fig.\ref{Refl_temps}, thus excluding twinning of the surface.)  A recent report on TbMnO$_{3}$ by Pimenov {\it et al.\/}\cite{PRL.102.107203} details the observation of a magnon and an electromagnon, active along perpendicular directions, resonating at the same frequency.  This occurrence gives the illusion of an isotropic magnetic excitation and is worthy of consideration in Cu$_{3}$Bi(SeO$_{3}$)$_{2}$O$_{2}$Cl.  Strongly opposing this argument are the nearly equivalent oscillator strengths measured in any two orthogonal polarizations, as seen in Fig.~\ref{polar0T}.  Electromagnons typically have oscillator strengths that are at least one order of magnitude greater than traditional magnons, as noted in the comparison of oscillator strengths above.   

A second more plausible explanation is the occurrence of weak magnons at two orthogonal polarizations.  N.~Kida {\it et al.\/}\cite{PRB.78.104414} observed a similar phenomena in DyMnO$_{3}$, namely, weak excitations arising with the a.c.~magnetic field oriented along both the {\it a\/} and {\it c\/} crystal axes.  They supported their interpretation of two orthogonal magnons by inelastic neutron scattering experiments on a similar rare-earth maganite where it was reported that two magnon dispersions curves from orthogonal axes crossed \textbf{k} = 0 at the same energy.  In Cu$_{3}$Bi(SeO$_{3}$)$_{2}$O$_{2}$Cl, we suspect that magnon dispersion curves from the [100] and [010] directions cross \textbf{k} = 0 at 33.1~cm$^{-1}$ ($\sim$4 meV); however, inelastic neutron scattering measurements are needed to support our hypothesis.

\subsubsection{$H\parallel\hat{c}$ field dependence}

To analyze the $H\parallel\hat{c}$ spectra (i.e., H parallel to the easy axis), we will utilize our d.c.~susceptibility measurements as well as generalize the magnetic structure determined for Cu$_{3}$Bi(SeO$_{3}$)$_{2}$O$_{2}$Br to Cu$_{3}$Bi(SeO$_{3}$)$_{2}$O$_{2}$Cl.  At zero field, six distinct magnetic sublattices can be identified, which would presumably lead to six distinct magnon branches; however, due to the canted nature of the copper ions occupying the 4(c) sites, the exact number of branches could differ from six.  At zero field, we only resolved one magnetic excitation (33.1~cm$^{-1}$).  No clear signature of the metamagnetic transition can be identified when tracking this excitation at fields between 0 and 1~T.  In a 1~T field, Cu$_{3}$Bi(SeO$_{3}$)$_{2}$O$_{2}$Cl has already undergone a metamagnetic transition where magnetic moments on every second layer flip, resulting in  ferromagnetic interlayer and canted ferrimagnetic behavior overall.\cite{arXiv:1203.2782v1}  (It should be noted that the c axis remains the easy axis after the transition).  Ironically, although the metamagnetic transition effectively reduces the number of magnetic sublattices from six to three, we observe an additional magnetic excitation appearing at fields of 1~T and greater.  At this point, two logical questions arise: First, why would the 33.1~cm$^{-1}$ excitation present in the low field antiferromagnetically-ordered state persist smoothly through the transition state and into the high-field ferro or ferrimagnetically ordered state?  Second, why are more magnetic excitations present in the high field phase despite a reduction in the number of magnetic sublattices?

As to the first question, magnons resulting from ferromagnetic resonance have been observed at infrared frequencies; however, they typically resonate at much lower frequencies because they must extrapolate linearly to zero frequency at zero applied magnetic field.  It is therefore very unlikely to observe a ferromagnetic resonance at 33.1~cm$^{-1}$ in a 1~T external field.  But there is an exceptional case that was first observed by Jacobs in FeCl$_{2}$ apropos metamagnetic transitions.\cite{JAP.36.1197}  When the metamagnetic transition occurs in FeCl$_{2}$, which signals a transition from a two-sublattice antiferromagnet to a ferromagnet, the AFMR line disappears in favor of a ferromagnetic resonance line around the same frequency.  The high frequency of the ferromagnetic resonance line in FeCl$_{2}$ is explained by large anisotropy fields in the material, and theoretical calculations support its existence.\cite{JPSJ.26.261}  We suspect that a similar situation occurs in Cu$_{3}$Bi(SeO$_{3}$)$_{2}$O$_{2}$Cl that explains the smooth movement of the excitation at 33.1~cm$^{-1}$ from below to above the metamagnetic transition.  Likewise, we suspect that the magnetic excitation at 10.5 cm$^{-1}$ (1 T) also originates from an antiferromagnetic resonance line, which would possess a zero field resonance frequency of 9.5 cm$^{-1}$ (slightly below our measurable range).

Apropos the second question, that is to say, the presence of more excitations above the metamagnetic transition versus below it contradicts the reduction in magnetic sublattices from six to three.  We suspect that more excitations do exist beneath the metamagnetic transition that are either below our measurable frequency range or that are too weak for us to resolve (resolution 0.3 cm$^{-1}$).  We have thoroughly inspected the low frequency spectra as well as the 33.1~cm$^{-1}$ excitation between 0 and 1~T, and we observe no clear signature of new resonances or the splitting of the existing 33.1~cm$^{-1}$ excitation.  It is possible that the 10.5~cm$^{-1}$ excitation (at 1~T) splits beneath the metamagnetic transition; however, below 1~T experimental limitations prevented us from tracking the excitation.  Future studies involving a high-resolution, low-frequency source are needed to investigate further.  

In the interval 1--10~T we can estimate the effective $g$ factor from the slope of the observed resonance lines at 10.5 and 33.1~cm$^{-1}$ using the formula $\hbar\omega=g\mu_{B}H$.\cite{PR.76.743}  The slopes of the resonance lines at 10.5 and 33.1~cm$^{-1}$ correspond respectively to $g$ factors of 2.16 and 0.24.

\subsubsection{$H\perp\hat{c}$ field dependence}

Again, we use our d.c.~susceptibility measurements and the neutron scattering results on Cu$_{3}$Bi(SeO$_{3}$)$_{2}$O$_{2}$Br to conclude that antiferromagnetic interactions exist within the $ab$ plane at all measurable fields in the $H\perp\hat{c}$ geometry.  The theory of antiferromagnetic resonances for uniaxial or cubic antiferromagnetic crystals was worked out by Keffer and Kittel,\cite{PR.85.329} who showed that when the external field is oriented perpendicular to the easy axis, the two \textbf{k} = 0 magnon branches correspond to frequencies $\omega/\gamma\cong\pm[2H_{E}H_{A}+H_{0}^{2}]^{1/2}$.  In the previous equation, $\gamma=ge/2mc$ where $g$ is the spectroscopic splitting factor, and $H_{0}$, $H_{A}$, and $H_{E}$ represent the static, anisotropy, and exchange fields respectively.  Generalizing this theory, which was worked out for two sublattices, to a situation with more than two sublattices, we can see that the dispersion of the excitation we observe (inset of Fig.~\ref{proof_HperpC}b) is in qualitative agreement (i.e., both are non-linear) with the lower frequency branch predicted by the AFMR theory.  (Further experiments are needed to determine $H_{A}$ , $H_{E}$, and $g$ and subsequently fit the formula to the data to either validate or undermine our generalization.)  When there is no external field, the two absorptions are predicted to be degenerate.  However, in antiferromagnets with strong magnetic anisotropy, which is the case for Cu$_{3}$Bi(SeO$_{3}$)$_{2}$O$_{2}$Cl, the degeneracy of magnon branches in zero field is lifted (e.g., MnO,\cite{PR.129.1566} NiO,\cite{PR.129.1566} and NiF$_{2}$\cite{PR.138.A1769}).  Following our previous reasoning that the resonance observed in Cu$_{3}$Bi(SeO$_{3}$)$_{2}$O$_{2}$Cl, which moves to lower frequency with increasing field, is the lower frequency branch of the \textbf{k} = 0 AFMR, we conclude that the higher frequency branch becomes masked, thus unobservable, behind the strong phonon absorptions starting $\sim$40 cm$^{-1}$.  AFMR theory also predicts that when the lower frequency branch reaches zero the external field is able to rotate the spins, at the expense of the anisotropy field, away from the easy axis ($c$ axis) and into the $ab$ plane.\cite{Mag.Scatt.Solids}  Extrapolating the lower frequency branch towards zero, while taking into account the uncertainty in the data points, we estimate a zero frequency crossing somewhere between 18.5 and 20.1 T.  Our predicted field range at which the spins rotate away from the easy axis is slightly higher, but remains in rough agreement with the values predicted on the Cu$_{3}$Bi(SeO$_{3}$)$_{2}$O$_{2}$Br analogue.\cite{arXiv:1203.2782v1}  But unlike Cu$_{3}$Bi(SeO$_{3}$)$_{2}$O$_{2}$Br, we can not identify intermediate and hard axes because the movement of our observed mode does not seem to depend on the orientation of the external field within the $ab$ plane.

\section{Conclusions}
The novel geometrically-frustrated layered compound Cu$_{3}$Bi(SeO$_{3}$)$_{2}$O$_{2}$Cl has been characterized using infrared spectroscopy, powder x-ray diffraction, and d.c.~magnetic susceptibility measurements.  Far-infrared reflectance measurements have revealed 16 new infrared-active phonon modes below 115~K.  The plethora of new modes observed strongly suggest a rearrangement of atomic positions within the unit cell; however, our subsequent powder x-ray diffraction measurements are completely consistent with the same 300~K structure ({\it Pmmn\/}) existing at 85~K.  Preliminary Raman spectra taken at 300~K on crystal 1 have revealed five phonon modes at frequencies close to five of the new modes observed in the infrared below 115~K.\cite{D.J.Arenas}  The results suggest a loss of inversion symmetry below 115~K.  Upon further investigation we have identified two non-centrosymmetric orthorhombic space groups ($Pm2_1n$ and $P2_1mn$) that have the same allowed Bragg reflection peaks as the 300~K {\it Pmmn\/} structure.  Therefore we suspect that a subtle second-order transition from {\it Pmmn\/} to either $Pm2_1n$ or $P2_1mn$ occurs near 115~K that is below the resolution of our powder x-ray diffraction experiment.  We plan to perform more sensitive measurements (\textit{e.g.,} single crystal or neutron powder diffraction) to resolve the uncertainty of this issue.  

In addition, an isotropic magnetic excitation is observed at 33.1~cm$^{-1}$ at 5~K.  We have tentatively assigned the magnetic excitation to a magnon based on analysis of previously reported oscillator strengths of magnons and electromagnons.  The isotropic behavior of the excitation within the $ab$ plane is potentially due to the Brillouin zone center crossing of two magnon dispersion curves along orthogonal directions, but inelastic neutron scattering measurements are needed to investigate further the excitation's seemingly isotropic existence.  

The resonance frequency of the 33.1~cm$^{-1}$ excitation strongly depends on the orientation of the static magnetic field.  An anisotropic response to the orientation of a static magnetic field is also seen in d.c.~susceptibility measurements on Cu$_{3}$Bi(SeO$_{3}$)$_{2}$O$_{2}$Cl, as well as neutron diffraction measurements on the similar Cu$_{3}$Bi(SeO$_{3}$)$_{2}$O$_{2}$Br compound.  

When the external magnetic field is applied parallel to the $c$ axis ($H\parallel\hat{c}$), the resonant frequency of the 33.1~cm$^{-1}$ excitation increases linearly with increasing field.  For fields of 1~T and greater applied along the $c$ axis an additional linearly-increasing magnetic excitation is observed (10.5~cm$^{-1}$ at 1~T).

When the external magnetic field is applied perpendicular to the $c$ axis ($H\perp\hat{c}$), the resonant frequency of the 33.1~cm$^{-1}$ excitation decreases quadratically with increasing field.  The results are in agreement with the behavior of an antiferromagnetic resonance line in the presence of strong magnetic anisotropy.

\begin{acknowledgments}
We acknowledge valuable discussions with E. S. Knowles, M. W. Meisel, and R. P. S. M. Lobo.  This work was supported by DOE through grant DE-FG02-02ER45984 at UF and DE-AC02-98CH10886 at the NSLS.                    

\end{acknowledgments}


\begin{thebibliography}{35}

\bibitem{NatMater.6.13} S.-W. Cheong and M. Mostovoy, \text{Nat. Mater.} \textbf{6}, 13 (2007).

\bibitem{PRB.82.060402} V. S. Zapf, M. Kenzelmann, F. Wolff-Fabris, F. Balakirev, and Y. Chen, \text{Phys. Rev. B} \textbf{82}, 060402 (2010).

\bibitem{PRL.92.257201} T. Goto, T. Kimura, G. Lawes, A. P. Ramirez, and Y. Tokura, \text{Phys. Rev. Lett.} \textbf{92}, 257201 (2004).

\bibitem{JMC.11.1152} P. Millet, B. Bastide, V. Pashchenko, S. Gnatchenko, V. Gapon, Y. Ksari, and A. Stepanov, \text{J. Mater. Chem.} \textbf{11}, 1152 (2001).

\bibitem{arXiv:1203.2782v1} M. Pregelj, O. Zaharko, A. G\"{u}nther, A. Loidl, and V. Tsurkan, \text{Eprint} \text{arXiv:1203.2782v1}.

\bibitem{PRB.80.214417} Ch. Kant, J. Deisenhofer, T. Rudolf, F. Mayr, F. Schrettle, A. Loidl, V. Gnezdilov, D. Wulferding. P. Lemmens, and V. Tsurkan \text{Phys. Rev. B} \textbf{80}, 214417 (2009).

\bibitem{PRL.94.137202} A. B. Sushkov, O. Tchernyshyov, W. Ratcliff II, S. W. Cheong, and H. D. Drew, \text{Phys. Rev. Lett.} \textbf{94}, 137202 (2005).

\bibitem{PR.129.1566} A. J. Sievers and M. Tinkham, \text{Phys. Rev.} \textbf{129}, 1566 (1963).

\bibitem{NatPhys.2.97} A. Pimenov, A. A. Mukhin, V. Y. Ivanov, V. D. Travkin, A. M. Balbashov, and A. Loidl, \text{Nat. Phys.} \textbf{2}, 97 (2006).

\bibitem{Born&Wolf} M. Born and E. Wolf, \textit{Principles of optics} (Cambridge University Press, United Kingdom, 1999).

\bibitem{topas}Bruker AXS (2005): TOPAS V3: General Profile and Structure
Analysis Software for Powder Diffraction Data, userÕs manual,
Bruker AXS, Karlsruhe, Germany.  Topas-Academic is available at www.topas-academic.net.

\bibitem{PRB.53.11734} A. Zibold, H. L. Liu, S. W. Moore, J. M. Graybeal, and D. B. Tanner, \text{Phys. Rev. B} \textbf{53}, 11734 (1996).

\bibitem{Wooten} F. Wooten, \textit{Optical Properties of Solids} (Academic Press, New York, 1972).

\bibitem{SMODES} H. T. Stokes and D. M. Hatch, 1999 \text{SMODES}, www.physics.byu.edu/stokesh/isotropy.html.

\bibitem{PRB.82.214302} D. J. Arenas, L. V. Gasparov, W. Qiu, J. C. Nino, C. H. Patterson, and D. B. Tanner, \text{Phys. Rev. B} \textbf{82}, 214302 (2010).

\bibitem{D.J.Arenas} D. J. Arenas and K. H. Miller \text{(unpublished).}

\bibitem{Radaelli1} P.G. Radaelli, Y. Horibe, M.J. Gutmann, H. Ishibashi, C.H. Chen, R.M. Ibberson, Y. Koyama, Y.S. Hor, V. Kiryukhin, and S.W. Cheong, Nature \textbf{416}, 155-158 (2002).

\bibitem{PR.185.211} H. Z. Cummins, \text{Phys. Rep.} \textbf{185}, 211 (1990).

\bibitem{PRB.34.278} P. Echegut, F. Gervais, and N. E. Massa, \text{Phys. Rev. B} \textbf{34}, 278 (1986).

\bibitem{PRB.30.6039} P. Echegut, F. Gervais, and N. E. Massa, \text{Phys. Rev. B} \textbf{30}, 6039 (1984).

\bibitem{PRB.39.4457} W. K. Lee and H. Z. Cummins, \text{Phys. Rev. B} \textbf{39}, 4457 (1989).

\bibitem{PRB.81.054428} P. Rovillain, M. Cazayous, Y. Gillais, A. Sacuto, M.-A. Measson, and H. Sakata, \text{Phys. Rev. B} \textbf{81}, 054428 (2010).

\bibitem{PRL.98.137206} D. Senff, P. Link, K. Hradil, A. Hiess, L. P. Regnault, Y. Sidis, N. Aliouane, D. N. Argyriou, and M. Braden, \text{Phys. Rev. Lett.} \textbf{98}, 137206 (2007).

\bibitem{PRL.101.187201} Y. Takahashi, N. Kida, Y. Yamasaki, J. Fujioka, T. Arima, R. Shimano, S. Miyahara, M. Mochizuki, N. Furukawa, and Y. Tokura, \text{Phys. Rev. Lett.} \textbf{101}, 187201 (2008).

\bibitem{PR.138.A1769} P. L. Richards, \text{Phys. Rev.} \textbf{138}, A1769 (1965).

\bibitem{JAP.38.1496} K. Aring and A. J. Sievers, \text{J. Appl. Phys.} \textbf{38}, 1496 (1967).

\bibitem{PRB.47.5300} S. G. Kaplan, T. W. Noh, A. J. Sievers, S-W. Cheong, and Z. Fisk, \text{Phys. Rev. B} \textbf{47}, 5300 (1993).

\bibitem{PRB.78.104414} N. Kida, Y. Ikebe, Y. Takahashi, J. P. He, Y. Kaneko, Y. Yamasaki, R. Shimano, T. Arima, N. Nagaosa, and Y. Tokura, \text{Phys. Rev. B} \textbf{78}, 104414 (2008).

\bibitem{PRL.102.107203} A. Pimenov, A. Shuvaev, A. Loidl, F. Schrettle, A. A. Mukhin, V. D. Travkin, V. Yu. Ivanov, and A. M. Balbashov, \text{Phys. Rev. Lett.} \textbf{102}, 107203 (2009).

\bibitem{JAP.41.980} A. J. Sievers, \text{J. Appl. Phys.} \textbf{41}, 980 (1970).

\bibitem{JAP.36.1197} I. S. Jacobs, S. Roberts, and P. E. Lawrence, \text{J. Appl. Phys.} \textbf{36}, 1197 (1965).

\bibitem{JPSJ.26.261} R. Alben, \text{J. Phys. Soc. Japan} \textbf{26}, 261 (1969).

\bibitem{PR.76.743} Charles Kittel, \text{Phys. Rev.} \textbf{76}, 743 (1949).

\bibitem{PR.85.329} F. Keffer and C. Kittel, \text{Phys. Rev.} \textbf{85}, 329 (1952).

\bibitem{Mag.Scatt.Solids} M. G. Cottam and D. J. Lockwood, \textit{Light Scattering in Magnetic Solids} (Wiley-Interscience, New York, 1986).

\end{thebibliography}
\end{document}